\documentstyle[12pt]{article}
\def\thebibliography#1{\medskip\section*{\centering
References\\}\medskip\list
{\arabic{enumi}.}{\settowidth\labelwidth{#1}\leftmargin\labelwidth
\advance\leftmargin\labelsep\usecounter{enumi}}
\def\newblock{\hskip .11em plus .33em minus .07em}
\sloppy\clubpenalty4000\widowpenalty4000 \sfcode`\.=1000\relax}
\def\op#1{\mathop{\fam0 #1}\limits}

\def\op#1{\mathop{\fam0 #1}\limits}
\newcommand{\Id}{{\rm Id\,}}
\newcommand{\ben}{\begin{eqnarray}}
\newcommand{\een}{\end{eqnarray}}
\newcommand{\beq}{\begin{equation}}
\newcommand{\eeq}{\end{equation}}
 \newcommand{\be}{\begin{eqnarray*}}
\newcommand{\ee}{\end{eqnarray*}}
\newcommand{\bea}{\begin{eqalph}}
\newcommand{\eea}{\end{eqalph}}
\newcommand{\cL}{{\cal L}}
\newcommand{\cF}{{\cal F}}
\newcommand{\cD}{{\cal D}}
\newcommand{\cT}{{\cal T}}
\newcommand{\bL}{{\bf L}}
\newcommand{\R}{{\bf R}}

\newcommand{\al}{\alpha}
\newcommand{\bt}{\beta}
\newcommand{\kp}{\kappa}
\newcommand{\dl}{\delta}
\newcommand{\la}{\lambda}
\newcommand{\ap}{\approx}
\newcommand{\om}{\omega}
\newcommand{\Om}{\Omega}
\newcommand{\m}{\mu}
\newcommand{\n}{\nu}
\newcommand{\g}{\gamma}
\newcommand{\G}{\Gamma}
\newcommand{\e}{\epsilon}
\newcommand{\ve}{\varepsilon}
\newcommand{\f}{\phi}

\newcommand{\Si}{\Sigma}
\newcommand{\si}{\sigma}
\newcommand{\w}{\wedge}
\newcommand{\wt}{\widetilde}
\newcommand{\wh}{\widehat}
\newcommand{\ol}{\overline}
\newcommand{\ot}{\otimes}
\newcommand{\dr}{\partial}
\newcounter{eqalph}
\newcounter{equationa}

\newenvironment{eqalph}{\stepcounter{equation}
\setcounter{equationa}{\value{equation}}
\setcounter{equation}{0}

\begin{eqnarray}}{\end{eqnarray}
\setcounter{equation}{\value{equationa}}}
 
\begin{document}
\hbox{}

\vskip2cm

\begin{center}{\large\bf Energy-momentum in gauge gravitation theory}
\bigskip

{\sc G. Sardanashvily and K. Kirillov}
\medskip
 
Physics Faculty, Moscow State University, 117234 Moscow, Russia
\end{center}
\bigskip

\begin{abstract} 
Building on the first variational formula of the calculus of variations, one
can derive the energy-momentum conservation laws from the condition
of the Lie derivative of gravitation Lagrangians along vector fields
corresponding to generators of general covariant transformations to be equal to
zero. The goal is to construct these vector fields. In gauge gravitation
theory, the difficulty arises because of fermion fields. General covariant 
transformations fail to preserve the Dirac spin structure
$S_h\to X$ on a world manifold $X$ which is associated with a 
certain tetrad field $h$. We introduce the universal Dirac spin structure $S\to
\Si\to X$ such that, given a tetrad field $h:X\to \Si$, the restriction of $S$
to $h(X)$ is isomorphic to $S_h$. The canonical lift
of vector fields on $X$ onto $S$ is constructed. We discover the
corresponding stress-energy-momentum conservation law. The
gravitational model in the presence of a background spin structure also is
examined. 
\end{abstract}

\section{Introduction}

There are several approaches to discovering energy-momentum conservation
laws in gravitation theory.
Here we are concerned with the gauge gravitation model 
of classical fields represented conventionally by sections of bundles $Y\to X$
over a world manifold $X$. In the framework of this model, the first
variational formula of the calculus of variations can be utilized in
order to discover differential conservation laws. This formula provides the
canonical decomposition of the Lie derivative of  Lagrangians along 
vector fields of infinitesimal gauge transformations into the two
terms. The first one is expressed into the variational derivatives and,
therefore, vanishes on shell.  Other is the divergence
$d_\la\cT^\la$ of the corresponding symmetry flow $\cT$. If a Lagrangian is
gauge-invariant, its  Lie derivative is equal to zero and the
weak conservation law
\beq
0\ap d_\la\cT^\la\label{01}
\eeq
takes place. 
When gauge transformations involve diffeomorphisms of a world manifold, i. e.,
their generators are the lift of vector fields $\tau$ on $X$ onto $Y$,
we have the stress-energy-momentum (SEM)
conservation law (\ref{01}) \cite{fat,fer,cam1,sar96b,sar97}.

In the model of metric gravity without matter \cite{nov}, the SEM conservation
law (\ref{01}) occurs due to invariance of the Hilbert--Einstein
Lagrangian under general covariant transformations. It takes the form
\beq
0\ap d_\la\cT^\la, \qquad  \cT^\la
\ap d_\m U^{\la\m},\label{K1} 
\eeq
where
\be
U^{\la\m} =\frac{\sqrt{-g}}{2\kp}
( g^{\la\nu}\nabla_\nu\tau^\m -g^{\m\nu}\nabla_\nu\tau^\la)  
\ee
is the well-known Komar superpotential. By $\nabla$
are meant the covariant derivatives with respect to the
Levi-Civita connection of a metric $g$.  

The same conservation law has been discovered in the Palatini model 
whose Lagrangian is an arbitrary function of the scalar curvature of  
a torsionless connection on $X$ \cite{bor}.

The affine-metric gravitation theory with tensor matter fields also has
been considered \cite{cam2,sar96b,giach96}. In case of a Lagrangian
expressed into the curvature tensor $R^\al{}_{\nu\la\m}$ of  
a general linear connection $K$ on $X$,  the SEM conservation
law is brought into the form (\ref{K1}) where
\beq
U^{\la\m} =2\frac{\dr\cL_{\rm AM}}{\dr R^\al{}_{\nu\la\m}} (\dr_\nu
\tau^\al + K^\al{}_{\si\nu}\tau^\si)  \label{K3}
\eeq
is the generalized Komar superpotential. 

In gauge gravitation theory, the difficulty arises 
because of Dirac fermion fields. 
These fields admit the Lorentz gauge
transformations only and, therefore, they are described in a pair with
a certain  tetrad field $h$ on $X$ by sections of the
corresponding spinor bundle $S_h\to X$. Recall that tetrad fields
constitute the 2-fold covering of pseudo-Riemannian metrics on $X$. Since
active general covariant transformations alter metrics on
$X$, they do not preserve spinor bundles
$S_h\to X$.  We thus face the problem how to
construct the appropriate lift of vector fields on $X$ onto spinor fields in
order to discover the SEM conservation laws. The solutions may be 
the following.

(i) The "world" spinors are considered. They carry 
representation of the universal covering $\ol{SL}(4,\R)$ of the
group $SL(4,\R)$ \cite{heh}.

(ii) The total system of
fermion-gravitation pairs is described \cite{fat96,sar96b,sard92}. 
In the present work, these pairs are represented
by sections of the composite spinor bundle $S\to\Si\to X$ where $\Si\to X$ is
the bundle of tetrad fields on $X$
\cite{sard9,sard95}. We show that gauge theory on this bundle is reduced to
the affine metric-gravitation theory in the presence of fermion fields. We
construct the canonical lift of vector fields
$\tau$ on $X$ onto the spinor bundle $S$ and, as a result, come to the
energy-momentum conservation law (\ref{K1}) where
$U^{\m\la}$ is the generalized Komar superpotential (\ref{K3}) with
accuracy to the standard term which is related to particular choice of the
Lepagean equivalent of a Lagrangian (see (\ref{979})).

(iii)  Given a tetrad field $h$, nonvertical gauge transformations which keep
$h$ are considered.
As a result, we come to the affine-metric modification of
the relativistic theory of gravity  by A.Logunov \cite{log86,log88} where
independent dynamic variables are non-metric gravitational fields
 and world connections on $X$ in the presence of a background
pseudo-Riemannian metric $g$. The Lagrangian of this model, by construction,
is invariant under the above-mentioned gauge transformations, but not 
the general covariant transformations. As a consequence, the total
energy-momentum flow is the sum of the superpotential term and
the metric energy-momentum tensor. It
is conserved if $g$ is the Minkowski metric.

Troughout, $X$ is a locally compact paracompact oriented connected 4-manifold.
It is assumed to be noncompact  and  parallelizable in order
that a pseudo-Riemannian metric, a spinor structure and a causal space-time
structure can exist on $X$.

\section{Conservation laws}

Differential operators and the Lagrangian formalism on sections of bundles are
phrased conventionally in the terms of jet
manifolds \cite{pom,sard94,sard95,sau}. 

As a shorthand, one can say that the
$k$-order jet manifold $J^kY$ of a bundle $Y\to X$
comprises the equivalence classes
$j^k_xs$, $x\in X$, of sections $s$ of $Y$ identified by the first $k+1$ 
terms of their Taylor series at a point $x$. 
Given bundle coordinates $(x^\la,y^i)$
of $Y\to X$, the $k$-jet manifold $J^kY$ is provided with the coordinates 
$(x^\la,y^i,y^i_\la,y^i_{\la\m},\ldots,y^i_{\la_1\cdots\la_k})$ where
$$ 
(y^i,y_\la^i,\ldots)(j^k_xs)=(s^i(x),\dr_\la s^i(x),\ldots).
$$

We are concerned with the first order Lagrangian formalism
on the configuration space $J^1Y$ coordinatized by 
$(x^\la,y^i,y^i_\la)$  together with the affine transition functions 
$$
{y'}^i_\la = \frac{\dr x^\m}{\dr{x'}^\la}(\dr_\m +y^j_\m\dr_j)y'^i. 
$$
A first order Lagrangian on $J^1Y$ is defined to be a horizontal density 
\beq
L=\cL(x^\m,y^i,y^i_\m)\om, \quad \om=dx^0\w\cdots\w dx^3.\label{03}
\eeq

Differential conservation laws in classical field theory are derived from
the condition of Lagrangians to be invariant under 1-parameter groups of gauge
transformations.

By a gauge transformation is meant an isomorphism
$\Phi$ of  the bundle $\pi:Y\to X$ over a diffeomorphism $f$ of $X$ which sends
the fibres $\pi^{-1}(x)$ onto the fibres $\pi^{-1}(f(x))$.
Every 1-parameter group $G[\al]$ of isomorphisms of 
$Y\to X$ over diffeomorphisms $f[\al]$ of $X$ yields the complete vector field
\beq
u=u^\la(x^\m)\dr_\la +u^i(x^\m,y^j)\dr_i \label{02}
\eeq
which plays the role of the generator of $G[\al]$. This
vector field is projected onto the vector field
$\tau_u=u^\m\dr_\m$ on $X$ which is the generator of 
$f[\al]$. For instance, if $G[\al]$ is the group of 
vertical isomorphisms over $\Id_X$ (or simply over X), the vector field
(\ref{02}) is vertical one $u=u^i(y)\dr_i$.  Conversely,  one can 
think on an arbitrary projective vector field
$u$ (\ref{02}) on a bundle $Y$ as being the generator of a lokal 1-parameter
groupof local isomorphisms of $Y$.

Given such a group $G[\al]$, the Lie derivative
$\bL_uL$  of a Lagrangian $L$ along its generator $u$ is equal to zero iff $L$
is invariant under $G[\al]$.

To calculate the Lie derivative $\bL_uL$ of the Lagrangian (\ref{03})
along the vector field $u$ (\ref{02}), one utilizes the canonical lift
\beq
j^1u=u^\la\dr_\la + u^i\dr_i + (d_\la u^i
- y_\m^i\dr_\la u^\m)\dr_i^\la   \label{44}
\eeq
of $u$ onto $J^1Y$ where $d_\la =\dr_\la +y^i_\la\dr_i
+y^i_{\m\la}\dr_i^\m$ denote the total derivatives.  We have 
\beq
\bL_uL=[\dr_\la u^\la\cL +(u^\la\dr_\la+
u^i\dr_i +(d_\la u^i -y^i_\m\dr_\la u^\m)\dr^\la_i)\cL]\om. \label{04}
\eeq

The first variational formula provides the canonical decomposition of
the Lie derivative (\ref{04}) in accordance with the variational task. This
decomposition takes the coordinate form
\ben
&& \dr_\la u^\la\cL +[u^\la\dr_\la+
u^i\dr_i +(d_\la u^i -y^i_\m\dr_\la u^\m)\dr^\la_i]\cL\equiv \label{C30}\\
&& \qquad   (u^i-y^i_\m u^\m )(\dr_i-d_\la \dr^\la_i)\cL -
d_\la[\pi^\la_i(u^\m y^i_\m -u^i) -u^\la\cL] \nonumber
\een
where
\beq
\dl_i\cL=(\dr_i- d_\la \dr^\la_i)\cL  \label{305}
\eeq
are the components of the Euler-Lagrange operator and
\beq
\cT=\cT^\la\om_\la =[\pi^\la_i(u^\m y^i_\m-u^i )- u^\la\cL]\om_\la, \quad
\pi^\la_i=\dr^\la_i\cL, \quad \om_\la=\dr_\la\rfloor\om, \label{Q30}
\eeq
is the symmetry flow along the vector field $u$.

It should be emphasized that the symmetry flow $\cT$ in 
the first variational formula (\ref{C30}) is not defined uniquely. It has the
general form
\beq
\cT^\la  =[\pi^\la_i(u^\m y^i_\m-u^i )-u^\la\cL]
+d_\m[c_i^{\la\mu}(y^i_\nu u^\nu-u^i)] \label{979}
\eeq
where $c_i^{\la\mu}=-c_i^{\mu\la}$ are arbitrary skew-symmetric functions on
$Y$ which correpond to different Lepagean equivalents of the Lagrangian
$L$. The flow 
(\ref{Q30}) corresponds to choice of the Poincar\'e--Cartan form as the
Lepagean equivalent of $L$.

The first variational formula (\ref{C30}) on shell 
\beq
(\dr_i- d_\la\dr^\la_i)\cL=0 \label{006}
\eeq
 comes to the weak transformation law 
\beq
\dr_\la u^\la\cL +[u^\la\dr_\la+
u^i\dr_i +(d_\la u^i -y^i_\m\dr_\la u^\m)\dr^\la_i]\cL 
 \ap -d_\la[\pi^\la_i(u^\m y^i_\m -u^i)-u^\la\cL]. \label{J4}
\eeq
If the Lie derivative $\bL_uL$
(\ref{04}) vanishes, we obtain the conservation law 
\beq
0\ap d_\la[\pi^\la_i(u^\m y^i_\m-u^i )-u^\la\cL]. \label{K4} 
\eeq
On solutions $s(x)$ of the differential Euler-Lagrange equations
\beq
\dr_i\cL-(\dr_\la+\dr_\la s^j\dr_j
+\dr_\la\dr_\m s^j \dr^\m_j)\dr^\la_i\cL=0,\label{2.29}
\eeq
the
weak identity (\ref{K4}) comes to the differential conservation law
\be
0 \ap \frac{d}{dx^\la}(\pi^\la_i(u^\m \dr_\m s^i-u^i) -u^\la\cL ).  
\ee

\section{Conservation laws and  background fields} 

Gauge symmetries of Lagrangians are broken in the presence of background
fields.

We restrict ourselves to the direct product $Y_{\rm tot}=Y\times Y'$
of a bundle $Y$ whose sections are dynamic fields and a bundle
$Y'$ whose sections play the role of background fields.
Let $Y$ and $Y'$ be coordinatized by $(x^\la, y^i)$ and $(x^\la, y^A)$
respectively. A total Lagrangian $L$ of dynamic and background fields 
is set on the configuration space $J^1Y_{\rm tot}$.
Dynamic fields are assumed to live on shell (\ref{006}),
whereas the background fields take the background values
\beq
y^B=\f^B(x), \quad y^B_\la= \dr_\la\f^B(x).\label{l69}
\eeq

Let us consider projectable vector fields 
\beq
u=u^\la(x)\dr_\la + u^A(x^\mu,y^B)\dr_A + u^i(x^\mu,y^B, y^j)\dr_i
\label{l68} 
\eeq
on $Y_{\rm tot}\to X$ which are
projectable on $Y\times Y'\to Y'$, for gauge 
transformations of background fields do not depend on dynamic fields.
Substituting (\ref{l68}) into (\ref{C30}),
we find the first variational formula in the presence of background fields.
This formula on shell (\ref{006}) of dynamic fields results in the weak
identity
\ben
&&\dr_\la u^\la\cL +[u^\la\dr_\la+  u^A\dr_A +
u^i\dr_i +(d_\la u^A -y^A_\m\dr_\la u^\m)\dr^\la_A + \label{l67} \\
&&\quad (d_\la u^i -y^i_\m\dr_\la
u^\m)\dr^\la_i]\cL\ap -d_\la[\pi^\la_i(u^\m y^i_\m -u^i)
-u^\la\cL]+ \nonumber \\
&&\quad (u^A-y^A_\la u^\la)\dr_A\cL + \pi^\la_Ad_\la (u^A-y^A_\mu
u^\mu). \nonumber
\een

In practice, total Lagrangians, by construction, are invariant under gauge
transformations, and their Lie derivatives along the corresponding 
vector fields (\ref{l68}) are equal to zero. 
In this case, the identity (\ref{l67}) is reduced to the transformation law
\beq
0\ap -d_\la[\pi^\la_i(u^\m y^i_\m-u^i) -u^\la\cL]+ 
(u^A-y^A_\la u^\la)\dr_A\cL + \pi^\la_A d_\la (u^A-y^A_\mu u^\mu) \label{l70}
\eeq
where the two last terms characterize disturbance of the
conservation law (\ref{K4}) in the presence of background fields (\ref{l69}).

\section{SEM conservation laws}

It is readily observed that the transformation law (\ref{J4}) is linear in
a vector field $u$. Therefore, one can consider superposition of the 
transformation laws along different vector fields. 

  For instance, every 
vector field on $Y$ projected onto a vector field $\tau$ on $X$
is the sum of the lift  of $\tau$ onto $Y$ and of a
vertical vector field on $Y$. It follows that 
every transformation law (\ref{J4})
derived from the first variational formula (\ref{C30}) appears to be 
superposition of (i) the Noether type transformation law 
\beq
[u^i\dr_i +d_\la u^i \dr^\la_i]\cL \ap d_\la(\pi^\la_iu^i) \label{0024}
\eeq
  for the Noether flow 
\beq
\cT^\la = -\pi^\la_iu^i \label{013}
\eeq
along a vertical vector field and of (ii) the SEM
transformation law along the lift of a vector field $\tau$ on $X$ onto
$Y$ \cite{cam1,sar96b,sar97}.

A vector field $\tau$ on $X$ gives rise to a vector field on a bundle $Y\to
X$ only by means of a connection on $Y$.

We follow the definition of connections on a bundle as sections  
\beq
\G=dx^\la\ot(\dr_\la+\G^i_\la\dr_i) \label{09}
\eeq
of the affine jet bundle $J^1Y\to Y$. For instance, a linear connection $K$ on
the tangent bundle $TX$ of 
$X$ and the dual connection $K^*$ on the cotangent bundle $T^*X$ read
\beq
 K^\al_\la=-K^\al{}_{\nu\la}(x)\dot x^\nu,\qquad 
K^*_{\al\la}=K^\nu{}_{\al\la}(x)\dot x_\nu \label{08}
\eeq
where $\dot x^\nu$ and $\dot x_\nu$ are induced coordinates with
respect to the holonomic frames $\{\dr_\nu\}$ and $\{dx^\nu\}$ in $TX$
and $T^*X$ respectively. We shall call (\ref{08}) the world connections.
Difference of two connections $\G$ and $\G'$ (\ref{09}) 
is a soldering 1-form
$$
\G-\G'=(\G^i_\la-\G'^i_\la)dx^\la\ot\dr_i
$$
on $Y$ which takes its values into the vertical tangent bundle $VY$ of $Y$.
Every connection $\G$ yields the first order differential operator 
\beq
D:J^1Y\to T^*X\op\ot_Y VY,
\qquad D =(y^i_\la -\G^i_\la)dx^\la\ot\dr_i, \label{010}
\eeq
on $Y$ called the covariant differential with respect to $\G$.

Let
$\tau=\tau^\m\dr_\m$ be a vector field on $X$ and
\beq
\tau_\G=\tau\rfloor\G=\tau^\m (\dr_\m+\G^i_\m\dr_i) \label{011}
\eeq
its horizontal lift onto $Y$ by a connection $\G$ (\ref{09})
on $Y$. The weak identity (\ref{J4}) along the vector field (\ref{011}) 
takes the form
\ben
&& {\bf L}_{\tau_\G}L\ap
-d_\la (\tau^\m \cT_\G{}^\la{}_\m)\om \label{504}\\
&&\dr_\m\tau^\m\cL + [\tau^\m\dr_\m
+\tau^\m\G^i_\m\dr_i +(d_\la(\tau^\m\G^i_\m)
 -y^i_\m\dr_\la\tau^\m)\dr^\la_i]\cL \ap \nonumber \\
&& \qquad - d_\la
[\pi^\la_i\tau^\m( y^i_\m- \G^i_\m)-\dl^\la_\m\tau^\m\cL], \nonumber
\een
where 
\beq
\cT_\G{}^\la{}_\m  =\pi^\la_i(y_\m^i -\G^i_\m)-\dl^\la_\m\cL \label{012}
\eeq
is termed the SEM
tensor relative to the connection $\G$ \cite{fer,cam1,sar96b,got92}. 

To obtain SEM conservation laws, one may choose different connections
$\G$ for different  
vector fields $\tau$ on $X$ and for different solutions of the
Euler--Lagrange equations.  
The SEM flows $\cT_\G$ and $\cT_{\G'}$ relative to different connections $\G$
and $\G'$ differ from each other in the Noether flow (\ref{013}) along the
vertical vector field $u=\tau^\m(\G^i_\m-\G'^i_\m)$. 

If all vector fields $\tau$ on $X$  gives rise onto $Y$ by means of the
same connection
$\G$, the transformation law (\ref{504}) is equivalent to the
system of weak equalities
\beq
(\dr_\m+\G^i_\m\dr_i +d_\la\G^i_\m \dr^\la_i)\cL\ap 
-d_\la\cT_\G{}^\la{}_\m. \label{Q31} 
\eeq

  For instance, let us choose the trivial local connection $\G_0{}^i_\m=0$.
In this case, the identity (\ref{504}) recovers the well-known transformation
law 
\beq
\frac{\dr\cL}{\dr x^\m} +\frac{d}{dx^\la} \cT_0{}^\la{}_\m (s)\ap 0
\label{Q40}
\eeq
of the canonical energy-momentum tensor
$$
\cT_0{}^\la{}_\m (s)= \pi^\la_i\dr_\m s^i -\dl^\la_\m\cL
$$
which however fails to be a true tensor. At the same time, the
transformation law (\ref{Q40}) on solutions $s$ of the differential
Euler--Lagrange equations (\ref{2.29}) is well-defined. It issues from the
identity (\ref{504}) when, for every solution $s$, we choose the proper
connection $\G$ such that $s$ is an integral section of $\G$, i. e.,
$\G^i_\m\circ s=\dr_\m s^i$.
The transformation law (\ref{Q40}) does not contain the Noether flows
associated with vertical gauge transformations. Each transformation law
(\ref{Q31}) on solutions $s$ of the differential Euler--Lagrange equations
(\ref{2.29}) represents a superposition of (\ref{Q40}) and the transformation
law
\beq
(\G^i_\m\dr_i +d_\la\G^i_\m \dr^\la_i)\cL\ap 
d_\la(\pi^\la_i\G^i_\la). \label{030} 
\eeq
The latter looks locally like the Noether transformation law (\ref{0024})
where $u^i=\tau^\m\G^i_\m$, but the entity
$\tau^\m\pi^\la_i\G^i_\m$ is not well-behaved.
At the same time, if the Lie derivatives of a Lagrangian $L$ along 
all local vector field $\tau^\m\G^i_\m$ vanishes, one can equate the left-hand
side of (\ref{030}) to zero. 

Building on this fact, one can use 
invariance of Lagrangians under vertical gauge transformations, e. g.,
under internal symmetry transformations in order to simplify 
SEM conservation laws. 
If a Lagrangian is not invariant under internal gauge
transformations, the total Lie derivative does not vanish and the 
SEM flow fails to be conserved in general.

\section{General covariant transformations}

Generators of general covariant transformations exemplify the bundle lift  of 
a vector field  $\tau$ on $X$ by means of the proper connection.

Let $LX\to X$ be the principal bundle of oriented linear frames in 
the tangent spaces to a world manifold $X$. Its structure group is 
$GL_4=GL^+(4,\R).$ Let $T$ denote the bundles associated with $LX$. 
They are exemplified by tensor bundles
\beq
T=(\op\ot^mTX)\ot(\op\ot^kT^*X). \label{971}
\eeq

Recall that a structure group $G$ of a principal bundle
$\pi_P:P\to X$ acts canonically on $P$ on the right 
\beq
r_g: p \mapsto pg, \quad \pi_P(p)=\pi_P(pg), \quad p\in P, \quad g\in
G. \label{022}
\eeq
Accordingly, every $P$-associated bundle $Y\to X$ with a standard fibre
$V$ is isomorphic to the quotient $Y=(P\times V)/G$
with respect to identification of elements $(p,v)$ and $(pg,g^{-1}v)$ for all 
$g\in G$. Isomorphisms $\Phi$ of a principal bundle $P$, by definition, are
equivariant under the canonical action (\ref{022}), that is,
$r_g\circ\Phi=\Phi\circ r_g.$ They generate the corresponding isomorphisms 
\beq
\Phi_Y: (P\times V)/G\to  (\Phi(P)\times V)/G \label{024}
\eeq
of bundles $Y$ associated with $P$.

Turn to the principal frame bundle $LX$. With respect to holonomic frames
$\{\dr_\m\}$, every element $\{H_a\}$ of $LX$ takes the form 
 $H_a=S^\m{}_a\dr_\m$ where $S^\m{}_a$ are matrices of representation of
the group $GL_4$ in $\R^4$. They constitute the bundle coordinates 
\be
(x^\la, S^\m{}_a), \qquad S'^\m{}_a=\frac{\dr x'^\m}{\dr x^\la}S^\la{}_a, 
\ee
of $LX$. Relative to these coordinates, the canonical action  
(\ref{022}) of $GL_4$ on $LX$ reads $r_g: S^\m{}_a\mapsto S^\m{}_bg^b{}_a$.

The percularity of the frame bundle $LX\to X$ lies in the fact that
every diffeomorphism $f$ of $X$ gives rise canonically to the isomorphism
\beq
\wt f: (x^\la, S^\la{}_a)\mapsto (f^\la(x),\dr_\m f^\la S^\m{}_a) \label{025}
\eeq
of $LX$ and to the corresponding
isomorphisms (\ref{024}) of associated bundles $T$. These isomorphisms
are called the general covariant
transformations or the holonomic isomorphisms because 
they send holonomic frames onto holonomic frames. 
 
The lift (\ref{025}) implies the canonical lift $\wt\tau$ of every
vector field $\tau$ on $X$ onto the principal bundle $LX$ and
the associated bundles. This lift takes the form
\beq
\wt\tau = \tau^\m\dr_\m + [\dr_\nu\tau^{\al_1}\dot
x^{\nu\al_2\cdots\al_m}_{\bt_1\cdots\bt_k} + \ldots
-\dr_{\bt_1}\tau^\nu \dot x^{\al_1\cdots\al_m}_{\nu\bt_2\cdots\bt_k}
-\ldots]\frac{\dr}{\dr \dot
x^{\al_1\cdots\al_m}_{\bt_1\cdots\bt_k}} \label{l28}
\eeq
on the tensor bundle (\ref{971}) and, in particular, 
\beq
\wt\tau = \tau^\m\dr_\m +\dr_\nu\tau^\al\dot x^\nu\frac{\dr}{\dr\dot x^\al},
\quad  
\wt\tau = \tau^\m\dr_\m -\dr_\bt\tau^\nu\dot x_\nu\frac{\dr}{\dr\dot x_\bt}
\label{l27} 
\eeq
on the tangent bundle $TX$ and the cotangent bundle $T^*X$ respectively.

In fact, the canonical lift $\wt\tau$ (\ref{l28}) is the 
horizontal lift (\ref{011}) by
means of the world connection (\ref{08}) which meets $\tau$ as a horizontal
vector field $\dr_\m\tau^\bt =-K^\bt{}_{\la\m}\tau^\la$. 

One can construct the horizontal lift $\tau_K$ of a vector field
$\tau$ on $X$ onto $LX$
by means of any world connection $K$. Such a lift is the generator of
nonholonomic isomorphisms of $LX$. 
Transformations of this type are called into play in the framework of gauge
theories of the general linear group $GL_4$ \cite{heh}.
At the same time, the term $-y_\m^i\dr_\la u^\m\dr_i^\la$ in the jet lift 
(\ref{44}) of a vector field on a bundle $Y$
is exactly the canonical lift (\ref{l27}). This is the
reason why we should consider the canonical lift $\wt\tau$ of 
vector fields on $X$ 
onto the $LX$-associated bundles $T$, in particular, onto
the bundle $\Si_g\subset\op\vee^2TX $ of pseudo-Riemannian metrics on $X$.
Indeed, gravitation Lagrangians are not invariant under nonholonomic gauge
transformations in general. 

\section{Superpotential}

The main percularity of SEM conservation laws along generators of general
covariant transformations consists in the phenomenon that
the corresponding SEM flow is reduced to the sum of a superpotential term and 
a terms which displays itself only in the presence of a background metric
on $X$.

To illustrate clearly this phenomenon, we consider tensor fields. Let 
$T$ be a tensor bundle coordinatized by $(x^\la,
y^A)$ where the collective index
$A$ is employed. Given a vector field $\tau$ on $X$, its canonical lift
$\wt\tau$ (\ref{l28}) on $T$ reads 
\beq
\wt\tau =\tau^\la\dr_\la + u^A{}_\al^\bt\dr_\bt\tau^\al\dr_A. \label{Q37}
\eeq
In this case, the strong equality $\bL_{\wt\tau}L=0$ and the weak identity
(\ref{K4}) take the coordinate form
\beq
\dr_\al(\tau^\al\cL) + u^A{}_\al^\bt\dr_\bt\tau^\al\dr_A\cL +
d_\m(u^A{}_\al^\bt\dr_\bt\tau^\al)\pi_A^\m -
 y^A_\al\dr_\bt\tau^\al\pi_A^\bt =0, \label{C310}
\eeq
\beq
0\ap  d_\la [g\pi^\la_A(y^A_\al\tau^\al-u^A{}_\al^\bt\dr_\bt\tau^\al)
-\tau^\la\cL]. \label{Q36}
\eeq
Due to the arbitrariness of the functions $\tau^\al$, the equality
(\ref{C310}) is equivalent to the system of equalities
\bea
&& \dr_\la\cL=0, \\
&& \dl^\bt_\al\cL + u^A{}_\al^\bt\dl_A\cL + d_\m(u^A{}_\al^\bt\pi_A^\m) =
 y^A_\al\pi_A^\bt,\label{C311a}\\
&& u^A{}_\al^\bt\pi_A^\m +u^A{}_\al^\m\pi_A^\bt =0. \label{C311b}
\eea

 Substituting the relations (\ref{C311a}) and (\ref{C311b})
into the weak identity (\ref{Q36}), we get the conservation law
\beq
0\ap  d_\la [u^A{}_\al^\la\dl_A\cL\tau^\al +
 d_\m(u^A{}_\al^\la\pi_A^\m\tau^\al)]
\label{C313}
\eeq
where $\dl_A\cL$ are the variational derivatives (\ref{305}). 
A glance at the expression (\ref{C313}) shows that the conserved flow on shell
takes the form  
\beq
\cT=W+d_HU, \quad \cT^\la = (W^\la +d_\m U^{\m\la}), \quad 
U^{\la\m}=-U^{\m\la}, \label{Q42}
\eeq
where $W\ap 0$ and 
\beq
U^{\la\m}(\tau) = u^A{}_\al^\la\pi_A^\m\tau^\al. \label{972}
\eeq
is a superpotential. 
It is readily observed that the superpotential (\ref{972}) arises since the
lift $\wt\tau$ (\ref{Q37}) depends on the
derivatives of the components of the vector field $\tau$.
At the same time, dependence of the superpotential (\ref{K3}) on the vector
field $\tau$ reflects the fact that the SEM conservation law (\ref{Q42}) is
maintained under general covariant transformations.

Also in gauge theory, if vector fields (\ref{02}) depend on derivatives of the
parameters of gauge transformations, the conserved flow $\cT$ is brought into
the form (\ref{Q42}). For instance, the conserved flow $-\cT^\la$ (\ref{Q42})
in electromagnetic theory is the familiar electric current, the superpotential
$4\pi U^{\m\la}$ consists with the electromagnetic strength, and
the relations (\ref{Q42}) are exactly the Maxwell equations.

Let us consider now tensor fields in the presence of a background
pseudo-Riemannian metric on a world manifold $X$. The total
configuration space of this model is the jet manifold $J^1Y$ of the product
$Y=T\times
\Si_g$ coordinatized by $(x^\la, y^A, \si^{\m\nu})$. Given a vector field
$\tau$ on $X$, its canonical lift onto $Y$ reads
\beq
\wt\tau =\tau^\la\dr_\la + u^A{}_\al^\bt\dr_\bt\tau^\al\dr_A +
(\dr_\nu\tau^\al\si^{\nu\bt} +\dr_\nu\tau^\bt\si^{\nu\al})\dr_{\al\bt}.
\label{973}
\eeq
Let the total Lagrangian $L$ of tensor fields and a metric field be
invariant under general covatiant transformations. Its Lie derivative along
the vector field (\ref{973}) is equal to zero. Then, 
we have the system of strong equalities
\bea
&& (\dr_\nu\tau^\al\si^{\nu\bt}
+\dr_\nu\tau^\bt\si^{\nu\al})\dr_{\al\bt}\cL +
\dl^\bt_\al\cL + u^A{}_\al^\bt\dl_A\cL +
 d_\m(u^A{}_\al^\bt\pi_A^\m) =y^A_\al\pi_A^\bt,\label{974b}\\
&& u^A{}_\al^\bt\pi_A^\m +u^A{}_\al^\m\pi_A^\bt =0, \label{974c}
\eea
and the weak transformation law (\ref{l70}) which restricted to
$\si^{\m\nu}= g^{\m\nu}(x)$ takes the form
\beq
0\ap - d_\la [\pi^\la_A(y^A_\al\tau^\al-u^A{}_\al^\bt\dr_\bt\tau^\al)
-\tau^\la\cL] + (\dr_\nu\tau^\al g^{\nu\bt}
+\dr_\nu\tau^\bt g^{\nu\al} -\dr_\la g^{\al\bt}\tau^\la)\dr_{\al\bt}\cL.
\label{975}
\eeq

Substituting (\ref{974b}) and (\ref{974c}) into (\ref{975}), we get the
transformation law
\be
0\ap -d_\la[\tau^\m t^\la_\m\sqrt{\mid g\mid} + u^A{}_\al^\la\dl_A\cL\tau^\al
+ d_\m(u^A{}_\al^\la\pi^\m_A\tau^\al)] +\dr_\la\tau^\m t^\la_\m\sqrt{\mid
g\mid} +\tau^\m t^\la_\bt\sqrt{\mid g\mid}\{^\bt{}_{\m\al}\} 
\ee
where 
\beq
t^\m_\bt\sqrt{\mid g\mid}= 2g^{\m\nu}\dr_{\al\bt}\cL \label{977}
\eeq
is the metric energy-momentum tensor of tensor fields. A glance at this
transformation law expression shows that the SEM flow of tensor fields in the
presence of a background metric is the sum 
\be
\cT^\la= \tau^\m t^\la_\m\sqrt{\mid g\mid} +
d_\m(u^A{}_\al^\la\pi^\m_A\tau^\al) 
\ee
of the metric energy-momentum tensor $t^\la_\m$ and the superpotential 
(\ref{972}). The latter does not make any contribution into the differential
transformation law which takes the familiar form $\nabla_\la t^\la_\m \ap 0.$

At the same time, if a metric field is dynamic, the metric
energy-momentum tensor (\ref{977}) on shell comes to zero. Thus, we observe
that the local
 part of the SEM flow in gravitation models displays itself only if symmetry
under general covariant transformations is broken. This is the phenomenon of
"hidden" energy.

\section{Energy-momentum of affine-metric gravity}

As we show below, gauge gravitation theory is reduced to the
affine-metric gravitation model in the presence of fermion fields. Therefore,
let us consider briefly the SEM conservation laws in  the affine-metric
theory of gravity without matter where dynamic variables are a
pseudo-Riemannian metric and a world connection  on $X$
\cite{cam2,giach96}.  Being principal connections on $LX$, the world 
connections are represented by sections $K$ of the quotient bundle
\beq
C:=J^1LX/GL_4\to X. \label{015}
\eeq
This bundle is coordinatized by $(x^\la, k^\nu{}_{\al\la})$ so that 
$k^\nu{}_{\al\la}\circ K=K^\nu{}_{\al\la}$
are coefficients of the world connection $K$ (\ref{08}) \cite{sard94,sard95}.
The bundle $C$ (\ref{015}) is not associated with $LX$, but it is an
affine bundle modelled on the vector bundle associated with $LX$. There exists
the canonical lift
\be
\wt\tau = \tau^\m\dr_\m +[\dr_\nu\tau^\al k^\nu{}_{\bt\m} - \dr_\bt\tau^\nu
k^\al{}_{\nu\m} - \dr_\m\tau^\nu
k^\al{}_{\bt\nu} -\dr_{\bt\m}\tau^\al]\frac{\dr}{\dr k^\al{}_{\bt\m}}
\ee
of a vector field $\tau$ on $X$ onto $C$.

The total configuration space of 
affine-metric gravity is the jet manifold $J^1Y$ of the product 
$Y=\Si_g\op\times_XC$ coordinatized by 
$(x^\la, \si^{\al\bt}, k^\al{}_{\bt\la}).$
Given a vector field $\tau$ on $X$, its canonical lift onto this product reads
\ben
&&\wt\tau =\tau^\la\dr_\la + (\si^{\nu\bt}\dr_\nu\tau^\al
+\si^{\al\nu}\dr_\nu\tau^\bt)\frac{\dr}{\dr \si^{\al\bt}}+ \label{Q50}\\
&& \qquad [\dr_\nu\tau^\al k^\nu{}_{\bt\m} - \dr_\bt\tau^\nu
k^\al{}_{\nu\m} - \dr_\m\tau^\nu
k^\al{}_{\bt\nu} -\dr_{\bt\m}\tau^\al]\frac{\dr}{\dr k^\al{}_{\bt\m}}.
\nonumber 
\een
  For the sake of simplicity, let us utilize the compact notation
$$
\wt\tau =\tau^\la\dr_\la + (u^A{}_\al^\bt\dr_\bt\tau^\al
-u^A{}_\al^{\ve\bt}\dr_{\ve\bt}\tau^\al)\dr_A.
$$

We assume that a Lagrangian density $L_{\rm AM}$ of affine-metric
gravity depends on the curvature tensor
\beq
R^\al{}_{\bt\n\la}=k^\al{}_{\bt\la\n}-
k^\al{}_{\bt\n\la}+k^\al{}_{\ve\n}k^\ve{}_{\bt\la}-k^\al{}_{\ve\la}
k^\ve{}_{\bt\n}. \label{101}
\eeq
Then, there are the corresponding relations
\beq
\frac{\dr\cL_{\rm AM}}{\dr k^\al{}_{\bt\nu}}= 
\pi_\si{}^{\bt\nu\la}k^\si{}_{\al\la}
-\pi_\al{}^{\si\nu\la}k^\bt{}_{\si\la},\qquad
 \pi_\al{}^{\bt\nu\la}= -\pi_\al{}^{\bt\la\nu}. \label{K300}
\eeq

Let $\bL_{\wt\tau}L_{\rm AM}=0$. We get the weak conservation law
\beq
0\ap  d_\la[ \pi^\la_A(u^A{}_\al^\bt\dr_\bt\tau^\al
-u^A{}_\al^{\ve\bt}\dr_{\ve\bt}\tau^\al -y^A_\al\tau^\al) +\tau^\la\cL]
\label{K8} 
\eeq
where
\be
&& \pi^\la_A u^A{}_\al^{\ve\bt} =\pi_\al{}^{\ve\bt\la},\\
&& \pi^\ve_A u^A{}_\al^\bt = \pi_\al{}^{\g\mu\ve}k^\bt{}_{\g\m} -
 \pi_\si{}^{\bt\mu\ve}k^\si{}_{\al\m} - \pi_\si{}^{\g\bt\ve}k^\si{}_{\g\al}
= \dr_\al{}^{\bt\ve}\cL_{\rm AM} - \pi_\si{}^{\g\bt\ve}k^\si{}_{\g\al} 
\ee
and the strong equality
\beq
\dl^\bt_\al\cL_{\rm AM} + \sqrt{-g}T_\al^\bt + u^A{}_\al^\bt\dr_A\cL_{\rm AM} +
 d_\m(u^A{}_\al^\bt)\pi_A^\m = y^A_\al\pi_A^\bt \label{K9}
\eeq
where one can think of 
$$
\sqrt{-g}T_\al^\bt =2\si^{\bt\nu}\dr_{\nu\al}\cL_{\rm AM}
$$
as being the metric energy-momentum tensor of world connections.

Substituting $y^A_\al\pi_A^\bt$ from the expression (\ref{K9})
into the conservation law (\ref{K8}), we bring the latter into the form
\be
 0\ap -
d_\la[\sqrt{-g}T^\la_\al\tau^\al -\pi^\la_A(u^A{}_\al^\bt\dr_\bt\tau^\al
+u^A{}_\al^{\ve\bt}\dr_{\ve\bt}\tau^\al) + u^A{}_\al^\la\tau^\al\dr_A\cL_{\rm
AM} +\pi^\m_A d_\m(u^A{}_\al^\la)\tau^\al]. 
\ee
After separating the variational derivatives, we find that the SEM 
conservation law (\ref{K8}) of affine-metric gravity comes to
the superpotential form 
\ben
&& 0\ap - d_\la[2g^{\la\m}\tau^\al\dl_{\al\m}\cL_{\rm AM}
+(k^\la{}_{\g\m}\dl_\al{}^{\g\m}\cL_{\rm AM} -
 k^\si{}_{\al\m}\dl_\si{}^{\la\m}\cL_{\rm AM} -
k^\si{}_{\g\al}\dl_\si{}^{\g\la}\cL_{\rm AM})\tau^\al - \nonumber \\
&& \qquad \dl_\al{}^{\ve\la}\cL_{\rm AM}\dr_\ve\tau^\al
+d_\m(\dl_\al{}^{\la\m}\cL_{\rm AM})\tau^\al + d_\m U^{\la\mu}]. \label{K11}
\een
 where $U$ is the generalized Komar superpotential (\ref{K3}). 

Let us now consider the total system of
the affine-metric gravity and, e.g., a covector Proca field coordinatized by
$y_\m=\dot x_\m$. Proca fields can model the gauge potentials when internal
symmetries are ignored. In virtue of the relation (\ref{C311b}), the Lagrangian
$L_{\rm P}$ of Proca fields depends  on the strength
$$
\cF_{\mu\nu} = y_{\nu\mu} -y_{\mu\nu} - \Om^\si{}_{\nu\mu}y_\si
$$
where $\Om^\si{}_{\nu\mu}=k^\si{}_{\nu\mu} -k^\si{}_{\mu\nu}$ is the torsion
tensor. In this case, the superpotential term in
the energy-momentum flow of the Proca fields (see (\ref{972})) is eliminated
due to the additional contribution
$-d_\m (\dr_\al{}^{\la\m}\cL_{\rm P}\tau^\al).$
The total SEM conservation law of affine-metric gravity and Proca fields takes
the form (\ref{K11}) where $L_{\rm AM}$ is replaced with $L_{\rm AM} +L_{\rm
P}$,
but $U$ remains the generalized Komar superpotential (\ref{K3}). We 
observe below that also fermion fields do not contribute into the total SEM
flow because of interaction with a torsion.

\section{Gauge gravitation theory}

In gauge gravitation theory, a gravitational field appears as a 
Higgs field corresponding to spontaneous symmetry breaking
because of the Dirac fermion fields \cite{sard92}.

We describe Dirac fermions as follows \cite{cra,obukh,rodr93}.
Given a Minkowski space $M$, let
${\bf C}_{1,3}$ be the complex Clifford algebra generated by elements
of $M$. A spinor space $V$ is defined to be a
minimal left ideal of ${\bf C}_{1,3}$  on
which this algebra acts on the left. We have representation
$\g: M\ot V \to V$ of elements of the Minkowski space $M\subset{\bf C}_{1,3}$
by the Dirac's matrices $\g$ on $V$.
 Let us consider the pairs $(l,l_s)$ of the Lorentz transformations $l$ of the
Minkowski space $M$ and the invertible elements $l_s$ of ${\bf C}_{1,3}$ such
that
$$
\g (lM\ot l_sV)=l_s\g (M\ot V).
$$
Elements $l_s$ constitute the Clifford group whose action on $M$
however is not effective. We take its spin
subgroup $L_s =SL(2,{\bf C})$ with the generators 
$$
I_{ab}=\frac{1}{4}[\g_a,\g_b].
$$
 
Let us consider some bundle of Clifford algebras ${\bf C}_{3,1}$
over a world manifold $X$. Its subbundles are both a spinor bundle $S_M\to X$
and the bundle $Y_M\to X$ of Minkowski spaces of generating elements of 
${\bf C}_{3,1}$.
To describe Dirac fermion fields on a world manifold $X$, the bundle
$Y_M$ must be isomorphic to the cotangent bundle $T^*X$
of $X$. It takes place if the structure group of
$LX$ is reducible to the connected Lorentz group $L$ and $LX$
contains a reduced $L$-principal subbundle $L^hX$ such that
$$
Y_M=(L^hX\times M)/L.
$$
In this case, the spinor bundle 
\beq
S_M=S_h=(P_h\times V)/L_s\label{510}
\eeq
is associated with the $L_s$-lift $P_h\op\to^z L^hX$ of $L^hX$.

In accordance with the well-known theorem, there is
the 1:1 correspondence between the reduced $L$ subbundles $L^hX$ of
$LX$ and the global sections $h$ of the quotient bundle
\beq
\Si:=LX/L\to X \label{5.15}
\eeq
with the standard fibre $GL_4/L$.
This bundle is the 2-fold covering of the bundle $\Si_g$ of pseudo-Riemannian
metrics. Its sections are called the tetrad fields. 
  
Every tetrad field $h$ defines an associated Lorentz atlas $\Psi^h$ of
$LX$ such that the corresponding local sections $z_\xi^h$ of $LX$
take their values into $L^hX$. Given $\Psi^h$ and a
holonomic atlas $\Psi=\{\psi_\xi\}$ of $LX$, the tetrad field $h$
can be represented by a family of local $GL_4$-valued tetrad functions
\beq
h_\xi=\psi_\xi\circ z^h_\xi,\qquad
dx^\la= h^\la_a(x)h^a, \qquad g^{\m\nu}=h^\m_ah^\nu_b\eta^{ab}. \label{L6}
\eeq
 
  For every tetrad field $h$, there exists  representation
\beq
\g_h: T^*X\ot S_h=(P_h\times (M\ot V))/L_s\to (P_h\times \g(M\ot V))/L_s=S_h
\label{L4}
\eeq
of covectors to a world manifold $X^4$ by the Dirac's $\g$-matrices
on elements of the spinor bundle $S_h$. Relative to an atlas
$\{z_\xi\}$ of $P_h$ and the associated Lorentz atlas $\{z^h_\xi=z\circ
z_\xi\}$ of $LX$, the representation (\ref{L4}) reads
$$
\g_h(h^a\ot y^Av_A(x))=\g^{aA}{}_By^Bv_A(x)
$$
where $\{v_A(x)\}$ are the corresponding fibre bases for $S_h$. As a shorthand,
we write
$$
\wh dx^\la=\g_h(dx^\la)=h^\la_a(x)\g^a.
$$
 
One may say that the representation (\ref{L4}) sets a Dirac spin structure
on a world manifold. Sections $\psi_h$ of the spinor bundle $S_h$ describe
Dirac fermion fields in the presence of the tetrad field $h$. Indeed,
let $A_h$ be a principal connection on $S_h$ and
\be
&&D: J^1S_h\to T^*X\ot S_h,\\
&&D=(y^A_\la-A^{ab}{}_\la I_{ab}{}^A{}_By^B)dx^\la\ot\dr_A,
\ee
the corresponding covariant differential (\ref{010}). Then, the first order
differential Dirac operator
\beq
\cD_h=\g_h\circ D: J^1S_h\to T^*X\ot S_h\to S_h, \label{I13}
\eeq
$$
 y^A\circ\cD_h=h^\la_a\g^{aA}{}_B(y^B_\la-A^{ab}{}_\la I_{ab}{}^A{}_By^B),
$$
is defined on $S_h$.

The crucial point lies in the fact that, for different  tetrad fields
$h$ and
$h'$,  the representations $\g_h$ and $\g_{h'}$
(\ref{L4}) are not equivalent.
Given two reduced Lorentz subbundles $L^hX$ and $L^{h'}X$, there exists the
vertical isomorphism $\Phi$ of the bundle $LX$ which sends
$L^hX$ onto $L^{h'}X$. However, it does not gives rise to a
morphism $\Phi_s:P_h\to P_{h'}$ in general. Moreover, the bundles  $L^hX$ and 
$L^{h'}X$ (accordingly, $S_h$ and $S_{h'}$) are not the equivalent locally
trivial bundles since the union of their atlases fails to be an atlas with the
Lorentz transition functions. This is the fact that characterizes
the symmetry breaking picture in gravitation theory and the physical nature of
gravity as a Higgs field. 

We may follow the
general procedure of describing spontaneous symmetry breaking in gauge theory
\cite{sard9,sard95}.
Since different tetrad fields $h$ and $h'$ define nonequivalent 
representations $\gamma_h$ and $\gamma_{h'}$, 
each Dirac fermion field is regarded only in a pair with
a certain tetrad field $h$. There is the 1:1 correspondence
between these pairs $(\psi_h,h)$ and the sections of the composite spinor
bundle 
\beq
S\to\Si\to X \label{L1}
\eeq
where $\Si$ is the quotient bundle (\ref{5.15}). This spinor bundle is defined
as follows \cite{sard92,sard95}.
 
Given the trivial L-principal bundle $LX\to\Si$, 
let us consider the $L_s$-principal bundle $P_\Si\to\Si$ which is the 2-fold
covering $P_\Si\op\to^z LX$ of $LX$.
In particular, there is imbedding of the $L_s$-lift $P_h$ of $L^hX$ 
onto the restriction of $P_\Si$ to $h(X)$.
Remark that the bundle $P_\Si\to X$ is not 
diffeomorphic to the universal covering of $LX$ with the structure group
$\ol{GL}(4,\R)$.
 
The composite spinor bundle $S\to\Si$ (\ref{L1}) is defined to be the $P_\Si$
associated bundle
$$
S= (P_\Si\times V)/L_s.
$$
One may say that the
bundle (\ref{L1}) introduces the universal Dirac spin structure on a world
manifold since, given a global section
$h$ of $\Si$, the restriction of $S_\Si$ to
$h(X)$ is somorphic to the spinor bundle $S_h$ (\ref{510}).
 
Let us provide the principal bundle $LX$ with a holonomic atlas
$\{\psi_\xi, U_\xi\}$ and the principal bundles $P_\Si$ and $LX\to\Si$
with associated atlases $\{z^s_\e, U_\e\}$ and $\{z_\e=z\circ z^s_\e\}$.
Relative to these atlases, the composite spinor bundle $S$ is endowed
with the bundle coordinates $(x^\la,\si_a^\m, \psi^A)$ where $(x^\la,
\si_a^\m)$ are coordinates of $\Si$ such that
$\si^\m_a$ are the matrix components of the group element
$(\psi_\xi\circ z_\e)(\si),$
$\si\in U_\e,\, \pi_{\Si X}(\si)\in U_\xi.$
Given a section $h$ of $\Si$, we have
$(\si^\la_a\circ h)(x)= h^\la_a(x)$
where $h^\la_a(x)$ are the tetrad functions (\ref{L6}).
 
Let us consider the bundle of Minkowski spaces
\[(LX\times M)/L\to\Si\]
associated with the $L$-principal bundle $LX\to\Si$. Since it is isomorphic
to the pullback $\Si\op\times_X T^*X$, there exists the representation 
\beq
\g_\Si: T^*X\op\ot_\Si S= (P_\Si\times (M\ot V))/L_s
\op\to_\Si (P_\Si\times\g(M\ot V))/L_s=S, \label{L7}
\eeq
\[\wh dx^\la=\g_\Si (dx^\la) =\si^\la_a\g^a.\]
This representation restricted to $h(X)\subset \Si$ recovers the morphism
$\g_h$ (\ref{L4}).
We are based on this representation in order to construct the total Dirac
operator on the composite spinor bundle $S$ (\ref{L1}) as follows. 

Every principal connection
\beq
\wt A=dx^\la\ot (\dr_\la +\wt A^B_\la\dr_B) + d\si^\m_a\ot
(\dr^a_\m+A^B{}^a_\m\dr_B) \label{100}
\eeq
on the bundle $S\to\Si$ yields the first order differential operator
\ben 
&&\wt D:J^1Y\to T^*X\op\ot_\Si S,\nonumber\\
&&\wt D=dx^\la\ot(\psi^B_\la-\wt A^B_\la -
A^B{}^a_\m\si^\m_{a\la})\dr_B \label{7.10}
\een
on $S$ \cite{sard9,sard95}. Let $h$ be a global section
of $\Si$ and $S_h$ the restriction of the bundle $S\to \Si$ to $h(X)$. The
corresponding restriction of $\wt D$ to $J^1S_h\subset S^1Y$
recovers to the familiar covariant differential on $S_h$ with respect 
to the connection
\beq
A_h=dx^\la\ot[\dr_\la+(\wt A^B_\la+A^B{}^a_\m\dr_\la h^\m_a)\dr_B]. \label{978}
\eeq
Building on the representation (\ref{L7}) and the
differential (\ref{7.10}), we construct the first order differential operator
\[\cD=\g_\Si\circ\wt D:J^1S\to T^*X\op\ot_\Si S\to S,\]
\[\psi^A\circ\cD=\si^\la_a\g^{aA}{}_B(\psi^B_\la-\wt A^B_\la -
A^B{}^a_\m\si^\m_{a\la}),\] on $S$.
One can think of $\cD$ as being the total Dirac operator since, for every 
tetrad field $h$, the restriction of $\cD$ to $J^1S_h\subset J^1S$ comes
to the Dirac operator $\cD_h$ (\ref{I13})
relative to the connection (\ref{978}) on the bundle $S_h$.

Let us choose the principal connection $\wt A$ (\ref{100}) on 
$P_\Si$ given by the following coefficients of the local connection form:
\beq
\wt A = (\wt A^{ab}{}_\m dx^\m+ A^{ab}{}^c_\m d\si^\m_c)\ot I_{ab},
\quad \wt A^{ab}{}_\m=\frac12 K^\n{}_{\la\m}\si^\la_c (\eta^{ca}\si^b_\n
-\eta^{cb}\si^a_\n ), \label{L10}
\eeq
\beq
A^{ab}{}^c_\m=\frac12(\eta^{ca}\si^b_\m -\eta^{cb}\si^a_\m), \label{M4}
\eeq
where  $K$ is a world connection on $X$ and (\ref{M4})
corresponds to the canonical left-invariant free-curvature connection on
the bundle $GL_4\to GL_4/L$. The corresponding differential $\wt D$
(\ref{7.10}) reads
\beq
\wt D =dx^\la\ot[\dr_\la -\frac12A^{ab}{}_\m^c (\si^\m_{c\la} +
K^\m{}_{\nu\la}  \si^\nu_c)I_{ab}{}^A{}_B\psi^B\dr_A]. \label{K104}
\eeq
Given a tetrad field $h$, the connection $\wt A$ (\ref{L10}) is reduced to the
principal connection  
$$
\wt K^{ab}{}_\la = A^{ab}{}_\m^c (\dr_\la h^\m_c + K^\m{}_{\nu\la}  h^\nu_c) 
$$
on $S_h$,  and the differential
(\ref{K104}) comes to the covariant differential of Dirac fermion fields in the
presence of the tetrad field $h$ and the world connection $K$
\cite{ar,pon,tuc}.  As a result, gauge
gravitation theory is reduced to the model of affine-metric
gravity and fermion fields. 

The total configuration space of this model is the jet manifold
$J^1Y$ of the bundle
\beq
Y=C\op\times_\Si S\op\oplus_\Si S^+ \label{042}
\eeq
coordinatized by $(x^\m,\si^\m_a, k^\m{}_{\nu\la},\psi^A,\psi^+_A)$. The total
Lagrangian on this configuration space is the sum
\beq
L=L_{\rm AM} + L_\psi \label{060}
\eeq
of (i) the affine-metric Lagrangian $L_{\rm
AM}(R^\al{}_{\bt\n\la},\si^{\m\n})$ expressed into the curvature tensor
$R^\al{}_{\bt\n\la}$ (\ref{101})
and the metric tensor $\si^{\m\n}=\si^\m_a\si^\nu_b\eta^{ab}$ and (ii) the
Lagrangian of fermion fields
\ben
&&L_\psi=\{\frac{i}2[ \psi^+_A(\g^0\g^\la)^A{}_B( \psi^B_\la -
\frac12A^{ab}{}_\m^c (\si^\m_{c\la} +
k^\m{}_{\nu\la}  \si^\nu_c)I_{ab}{}^B{}_C\psi^C) -\label{230}  \\
&& \qquad ( \psi^+_{A\la}- \frac12A^{ab}{}_\m^c (\si^\m_{c\la} +
k^\m{}_{\nu\la}  \si^\nu_c)\psi^+_C
I^+_{ab}{}^C{}_A)(\g^0\g^\la)^A{}_B\psi^B] -\nonumber\\
&& \qquad m\psi^+_A(\g^0)^A{}_B\psi^B\}\sqrt{\mid\si\mid}\om \nonumber
\een
where $\g^\m=\si^\m_a\g^a$ and $\si=\det(\si_\m\nu)$. One can easily verify
that
\beq
\frac{\dr\cL_\psi}{\dr k^\m{}_{\nu\la}} + 
\frac{\dr\cL_\psi}{\dr k^\m{}_{\la\nu}} =0, \label{2C14}
\eeq
that is, the Lagrangian (\ref{230}) depends on the torsion of world connections
$k$. 

\section{Energy-momentum of gauge gravity}

Recall: (i) that there is the 1:1 correspondence 
between the invariant vector fields on the groups $L$ and $L_s$ and 
(ii) that, for every connection form on $LX$, its Lorentz-valued component
restricted to $L^hX$ is a connection on
$L^hX$. Building on these facts, we can define the canonical lift
\ben
&&\wt\tau = \tau^\m\dr_\m +
(\dr_\nu\tau^\al k^\nu{}_{\bt\m} - \dr_\bt\tau^\nu
k^\al{}_{\nu\m} - \dr_\m\tau^\nu
k^\al{}_{\bt\nu} -\dr_{\bt\m}\tau^\al)\frac{\dr}{\dr k^\al{}_{\bt\m}}+
\label{K105}\\
&& \qquad \dr_\nu\tau^\m \si^\nu_c \frac{\dr}{\dr \si^\m_c} +
 \frac12\dr_\al\tau^\bt(\eta^{ap}\si^\al_p\si^b_\bt
-\eta^{bp}\si^\al_p\si^a_\bt)
[(\eta_{bc}\dl_a^d - \eta_{ac}\dl_b^d)\si^\m_d \frac{\dr}{\dr \si^\m_c} +
\nonumber \\
&& \qquad  I_{ab}{}^A{}_B\psi^B\dr_A+ I^+_{ab}{}^A{}_B\psi^+_A\dr^B]. \nonumber
\een
of a vector field $\tau$ on $X$ onto the bundle (\ref{042}). For the sake of
simplicity, let us employ again the compact notation
$$
\wt\tau =\tau^\m\dr_\m + \dr_\nu\tau^\m \si^\nu_a\frac{\dr}{\dr \si^\m_a} +
(u^A{}_\al^\bt\dr_\bt\tau^\al -u^A{}_\al^{\ve\bt}\dr_{\ve\bt}\tau^\al)\dr_A.
$$ 

The Lagrangian (\ref{060}), by construction, is invariant 
under transformations of the holonomic atlases of the principal frame
bundle $LX$ (passive general covariant transformations acting on the Greek
indices) and under transformations of the atlases of the principal bundle
$P_\Si$ and $LX\to\Si$ (passive Lorentz gauge transformations acting on
the Latin indices). It follows that this Lagrangian is invariant under
infinitesimal active gauge transformations 
whose generator is the canonical lift (\ref{K105}) and 
\beq
\bL_{\wt\tau}L=0. \label{K200}
\eeq
Then, the weak conservation law
\ben
&&0\ap d_\la[ \dr^\la_A\cL_{\rm AM}(u^A{}_\al^\bt\dr_\bt\tau^\al
-u^A{}_\al^{\ve\bt}\dr_{\ve\bt}\tau^\al -y^A_\al\tau^\al) \nonumber\\
&& \qquad +\frac{\dr\cL_\psi}{\dr\si^\al_{c\la}} (\dr_\bt\tau^\al\si^\bt_c -
\si^\al_{c\m}\tau^\m) - \frac{\dr\cL_\psi}{\dr\psi^A_\la}\psi^A_\al\tau^\al -
\frac{\dr\cL_\psi}{\dr\psi^+_{A\la}}\psi^+_{A\al}\tau^\al
+\tau^\la\cL] \label{K400}
\een
takes place. We have also the relations (\ref{K300}) and the relation 
$$
\frac{\dr\cL_\psi}{\dr k^\m_{\nu\la}} =\frac{\dr\cL_\psi}{\dr\si^\m_{c\la}}
\si^\nu_c.
$$

Due to arbitrariness of the functions $\tau^\al$, the equality
(\ref{K200}) implies the strong equality (\ref{K9}) where $\sqrt{\mid g\mid}$
is replaced by $\sqrt{\mid\si\mid}$ and the strong equality
\beq
\dl_\al^\bt\cL_\psi +\sqrt{\mid\si\mid} t_\al^\bt +
\frac{\dr\cL_\psi}{\dr\si^\al_{c\la}}
\si^\bt_{c\la} - \frac{\dr\cL_\psi}{\dr\si^\m_{c\bt}}\si^\m_{c\al} +
\dr_A\cL_\psi u^{A\bt}_\al - \frac{\dr\cL_\psi}{\dr\psi^A_\bt}\psi^A_\al -
\frac{\dr\cL_\psi}{\dr\psi^+_{A\bt}}\psi^+_{A\al}=0 \label{K301}
\eeq
where 
$$
\sqrt{\mid\si\mid} t^\bt_\al =\si^\bt_a\frac{\dr\cL_\psi}{\dr\si^\al_a}.
$$

Substituting the term $y^A_\al\dr_A^\bt\cL_{\rm AM}$  from the expression
(\ref{K9}) and the term
$$
\frac{\dr\cL_\psi}{\dr\si^\m_{c\bt}}\si^\m_{c\al} + 
\frac{\dr\cL_\psi}{\dr\psi^A_\bt}\psi^A_\al +
\frac{\dr\cL_\psi}{\dr\psi^+_{A\bt}}\psi^+_{A\al}
$$
  from the expression (\ref{K301}) 
into the conservation law (\ref{K400}), we bring the latter into the form
\ben
&& 0\ap d_\la[-\si^\la_a\tau^\al\dl^a_\al\cL
-(k^\la{}_{\g\m}\dl_\al{}^{\g\m}\cL_{\rm AM} -
 k^\si{}_{\al\m}\dl_\si{}^{\la\m}\cL_{\rm AM} -
k^\si{}_{\g\al}\dl_\si{}^{\g\la}\cL_{\rm AM})\tau^\al + \nonumber\\
&& \quad
\dl_\al{}^{\ve\la}\cL_{\rm AM}\dr_\ve\tau^\al
-d_\m(\dl_\al{}^{\la\m}\cL_{\rm AM})\tau^\al]
- d_\la[d_\m(\pi_\al{}^{\nu\m\la}(D_\nu\tau^\al+
\Om^\al{}_{\nu\si}\tau^\si)] \nonumber\\ 
&&\quad + d_\la[(\frac{\dr\cL_\psi}{\dr\si^\al_{a\m}}\si^\la_a +
\frac{\dr\cL_\psi}{\dr\si^\al_{a\la}}\si^\m_a)\dr_\m\tau^\al]. \label{K303}
 \een
In accordance with the relation (\ref{2C14}), the last term in the expression 
(\ref{K303}) is equal to zero, i. e., fermion fields do not contribute into the
superpotential. It follows that the SEM conservation law (\ref{K400}) comes to
the form (\ref{K1}) where $U$ is the generalized  Komar superpotential
(\ref{K3}). 

We thus may conclude that the generalized Komar superpotential (\ref{K3})
appears to be universal for several gravitation models.

\section{Relativistic theory of gravity}

There is another lift $\Phi$ of diffeomorphisms $f$ of the world manifold $X$
onto
$LX$  which preserve a given tetrad field $h$. In this case, we have 
\beq
\g_h(dx^\la)\to \g_h(\Phi^\la{}_\m dx^\m) = \Phi^\la{}_\m h^\m_a\g^a.
\label{ll4}
\eeq
It is readily observed that the Dirac operator (\ref{I13}) is not
invariant under these transformations. To overcome this difficulty, one may
introduce an additional geometric field $q$ represented 
locally by tensor functions
$q^\la{}_\m$ and then rewrite the Dirac operator into the form
\beq
\cD^{qh}= q^\la{}_\m h^\m_a\g^a D^h{}_\la  \label{ll5}
\eeq
which is invariant 
$$
\cD^{qh}\to \Phi^\la{}_\nu q^\la{}_\m h^\m_a\g^a D'^h{}_\la=
q'^\la{}_\m h^\m_a\g^a D'^h{}_\la
$$
under the transformations (\ref{ll4}).
As a result, we get the affine-metric modification of 
the relativistic theory of gravity (RTG) \cite{log86,log88}
where independent dynamic variables are non-metric gravitational fields $q$ and
world connections $\G$ on $X$ in the presence of a background tetrad (or
metric) field, e. g. the Minkowski metric. Since the above-mentioned gauge
transformations do not act on the indices of the background
Minkowski metric
$\eta^{\m\nu}$  and the low index $\m$ of a gravitational field
 $q^\la{}_\m$, it is the tensor field
\beq
q^{\m\nu}=q^\m{}_\al q^\nu{}_\bt g^{\al\bt}=H^\m_aH^\nu_b\eta^{ab} \label{06}
\eeq
which perform contraction of indices in the Lagrangians of matter fields and
world connections. Thus, the tensor field  (\ref{06}) plays the role of an
effective metric. Obviously, it is not a true  metric, for the coframes 
$H^a=H^a_\m dx^\m$ are carried by the $\g$-matrices
$$
\g_h(H^a)= h_{\m\nu}\eta^{ab}q^\nu{}_\al h^\al_bh^\m_c\g^c 
$$
with respect to the same representation as the coframes
$h^a=h^a_\m dx^\m$ are done. At the same time, there always exists 
a pseudo-Riemannian metric $g'$  such that $g'^{\m\nu}=q^{\m\nu}$.

Note that additional tensor fields $q$ and generalized
tetrad fields $H^\m_a=q^\m{}_\nu h^\nu_a$ have been considered in the
framework of both the
$GL_4$-gauge models
\cite{perc91,heh} and the gauge model of the translation group where $q$
describe deformations of a world manifold \cite{sard92}. 

Now let us describe our construction in details.
Given a tetrad field $h$, let $\G_h$ be a principal connection on $L^hX$
extended to $LX$. Any 1-parameter group 
$f[\al]$ of diffeomorphisms of $X$ whose generator is a vector field
$\tau$  on $X$ gives rise to the local group of nonholonomic isomorphisms 
$f_\G[\al]$ of  $LX$ whose generator is the vector field
\be
\tau_\G=\tau^\la(\dr_\la -\G_h{}^\m{}_{\nu\la}S^\nu{}_a\frac{\dr}{\dr
S^\m{}_a}) 
\ee
on $LX$. Isomorphisms $f_\G[\al]$ keep the Lorentz subbundle
$L^hX$ and the tetrad field $h$. Given a 1-parameter group
$\wt f[\al]$ of general covariant transformations over diffeomorphisms
$f[\al]$, its elements are represented by composition
\beq
\wt f[\al] =\Phi[\al]\circ f_\G[\al] \label{045}
\eeq
of $f_\G[\al]$ and the vertical isomorphisms $\Phi[\al]$ of $LX$ with
the generator 
$$
u=D_\nu\tau^\m S^\nu{}_a\frac{\dr}{\dr S^\m{}_a}
$$
where by $D_\nu$ are meant the covariant derivatives with respect to the
connection $\G_h$. 

Recall the following. Let $P\to X$ be a principal bundle with a
structure group $G$ and
$\wh P\to X$ the associated group bundle with the standard fibre $G$ on
which the structure group acts by the adjoint representation. Given a
bundle $Y$ associated with $P$, there is the canonical fibre-to-fibre
morphism $\wh P\times Y\to Y$. As a consequence, every vertical isomorphism
$\Phi_Y$ (\ref{024}) of an associated bundle 
$Y$ is brought into the form
$$
\Phi_Y(y)=q_\Phi(\pi(y))\cdot y, \qquad y\in Y, 
$$
where $q_\Phi(x)$ is a global section of the group bundle $\wt P$. In
particular, we can show that the matrix functions $\Phi^\la{}_\m(x)$ in the
expression (\ref{ll4}) are local functions of a section 
$q_{\Phi[\al]}$ of the group bundle
$\wh{LX}\subset TX\ot T^*X$ which corresponds to the vertical isomorphism 
$\Phi[\al]$ in the expression (\ref{045}). 

Therefore, let us introduce additional dynamic fields $q$ represented by
sections of the group bundle $\wh{LX}$ and consider the compositions $q\cdot
h$. Then, every vertical gauge transformation of $q\cdot h$ reads
$$
\Phi(q\cdot h) =q_\Phi\cdot q\cdot h=q'\cdot h.
$$
The field $q$ in the expression (\ref{ll5}) exemplifies such a section of
$\wh{LX}$.

Thus, we come to RTG in case of a background tetrad field $h$, the dynamic
gravitational fields $q$ and the effective tetrad fields
$H=q\cdot h$. The total configuration space of RTG in the 
presence of fermion fields is the jet manifold of the bundle
\beq
Y=\wh{LX}\op\times_X C\op\times_X S_h\op\oplus_X S_h^+\label{055}
\eeq
coordinatized by $(x^\m,q^\m{}_\nu, k^\m{}_{\nu\al},\psi^A,\psi^+_A)$.
Let us consider the gauge transformations of the bundle (\ref{055}) whose
generator is the vector field
\ben
&&\wt\tau = \tau^\m\dr_\m +
(\dr_\nu\tau^\al k^\nu{}_{\bt\m} - \dr_\bt\tau^\nu
k^\al{}_{\nu\m} - \dr_\m\tau^\nu
k^\al{}_{\bt\nu} -\dr_{\bt\m}\tau^\al)\frac{\dr}{\dr k^\al{}_{\bt\m}}+
\label{051}\\
&& \quad (\dr_\nu\tau^\m q^\nu{}_\al+ 
\tau^\la\G_h{}^\nu{}_{\al\la}q^\m{}_\nu)\frac{\dr}{\dr q^\m_\al}+ \nonumber\\
 &&\quad \frac12\dr_\al\tau^\bt(\eta^{ap}h^\al_ph^b_\bt
-\eta^{bp}h^\al_ph^a_\bt)[I_{ab}{}^A{}_B\psi^B\dr_A+
I^+_{ab}{}^A{}_B\psi^+_A\dr^B]. \nonumber
\een
These transformations act on the metric functions 
$g^{\m\nu}$ and the tetrad functions $h^\m_a$ of the background field $h$ by
the law
\be
&& g^{\m\nu} \to \tau^\la(\dr_\la g^{\m\nu} -\G_h{}^\m{}_{\al\la}g^{\al\nu}
-\G_h{}^\nu{}_{\al\la}g^{\m\al}),\\
&& h^\m_a \to \tau^\la(\dr_\la h^\m_a -\G_h{}^\m{}_{\al\la}h^\al_a) +
\dr_\al\tau^\bt(\eta^{dp}\eta_{ba}h^\al_ph^b_\bt-h^\al_a h^d_\bt)h^\m_d.
\ee
At the same time, they act on the effective metric field $q$ in
accordance with the standard law of general covariant transformations
\be
&& q^{\m\nu} \to \tau^\la\dr_\la q^{\m\nu} +
\dr_\al\tau^\m q^{\al\nu} + \dr_\al\tau^\nu q^{\m\al}, \\
&& H^\m_a \to \tau^\la\dr_\la H^\m_a + \dr_\nu\tau^\m H^\nu_a +
\dr_\al\tau^\bt(\eta^{dp}\eta_{ba}h^\al_ph^b_\bt -h^\al_a h^d_\bt)H^\m_d.
\ee
The total gauge-invariant Lagrangian of RTG can be given by the sum
\beq
L_{\rm RG}=L_{\rm AM}(q,\G) +L_\psi +L_q(q,g) \label{040}
\eeq
of the Lagrangian  $L_{\rm AM}$ (\ref{060}) of the affine-metric gravity
and the Lagrangian $L_\psi$ of fermion fields 
 where the metric field $\si$ is replaced by the effective metric
$q$ (\ref{06})  and  of a Lagrangian $L_q$ of fields $q$ where contraction is
performed by means of the background metric $g$. Therefore,
$L_q$ is not invariant under familiar general covariant transformations. 

In particular, put
\beq
L_{\rm AM}=(-\la_1R+\la_2)\sqrt{\mid g\mid}, \quad L_\psi=0, \quad
L_q=\la_3\eta_{\m\nu}g^{\m\nu}\sqrt{\mid\eta\mid} \label{039}
\eeq
where $R=g^{\m\nu}R^\al{}_{\m\al\nu}$ is the effective scalar curvature of a
connection $\G$. Then, the conventional RTG is recovered.

Let us examine the SEM conservation laws in the RTG model. Building on the
formula (\ref{l70}), we get 
$$
0\ap d_\la(2\tau^\m g^{\la\al}\frac{\dr\cL_{\rm RG}}{\dr g^{\al\m}} +
d_\m U^{\la\m})
+ (\tau^\la\dr_\la g^{\m\nu} -\dr_\al\tau^\m g^{\al\nu} +
\dr_\al\tau^\nu g^{\m\al})\frac{\dr\cL_{\rm RG}}{\dr g^{\m\nu}}
$$
where $U$ is the generalized Komar superpotential.
A glance at this relation shows that, if the Lagrangian $L_{\rm RG}$ contains
the Higgs term $L_q$, the energy-momentum flow does not reduce to the
superpotential, and we have the standard covariant conservation law
\beq
\nabla_\al t^\al_\la \ap 0, \qquad \sqrt{\mid g\mid}t^\al_\la=2g^{\al\m}
\frac{\dr\cL_q}{\dr g^{\m\nu}}. \label{061}
\eeq
 In particular, it
is brought into the conservation law if $g$ is the Minkowski
metric. 

Note that 
one can express the energy-momentum tensor $t$ 
(\ref{061}) into the tensor
$$
\sqrt{\mid g\mid} T^\al_\la=2q^{\al\m}
\frac{\dr\cL}{\dr q^{\m\nu}}
$$
owing to the field equations
\be
\frac{\dr\cL_{\rm RG}}{\dr q^{\al\m}}=\frac{\dr\cL}{\dr q^{\al\m}}+
\frac{\dr\cL_q}{\dr g^{\al\m}}=0. 
\ee
In particular, if $L_q$ is equal to the Higgs massive term
(\ref{039}), we have the relation
$$
\sqrt{\mid g\mid}t^\al_\la=\sqrt{\mid q\mid}(T^\al_\la+
\frac12\delta^\al_\la T^\m_\m).
$$
Let us emphasize that this relation does not depend on the constant
$\la_3$ which therefore can be as small as will.

\end{document}